\documentstyle[12pt]{article}
\newcommand{\nc}{\newcommand}
\nc{\al}{\alpha}
\nc{\ald}{{\dot \al}}
\nc{\ba}{\beta_\al}
\nc{\bb}{\beta_\beta}
\nc{\ga}{\g^\al}
\nc{\gb}{\g^\beta}
\nc{\db}{\pa_\beta}
\nc{\dtb}{\delta_\theta^\beta}
\nc{\dab}{{\delta_\al}^\beta}
\nc{\vmab}{V_{-\al}^\beta}
\nc{\vab}{V_\al^\beta}
\nc{\vib}{V_i^\beta}
\nc{\g}{\gamma}
\nc{\G}{\Gamma}
\nc{\D}{\Delta}
\nc{\paj}{P_{-\al}^j}
\nc{\la}{\lambda}
\nc{\La}{\Lambda}
\nc{\var}{\varphi}
\nc{\pa}{\partial}
\nc{\nn}{\nonumber \\ }
\nc{\hf}{\frac{1}{2}}         
\nc{\dz}{\frac{dz}{2\pi i}}
\nc{\fabc}{{f_{ab}}^c}
\nc{\binomial}[2]{\left (\begin{array}{c} {#1}\\ {#2} \end{array}
\right )}
\nc{\ben}{\begin{equation}}
\nc{\een}{\end{equation}}
\nc{\bea}{\begin{eqnarray}}
\nc{\eea}{\end{eqnarray}}
\nc{\bra}[1]{\langle {#1}|}
\nc{\ket}[1]{|{#1}\rangle}

\nc{\C}{\mbox{\hspace{1.24mm}\rule{0.2mm}{2.5mm}\hspace{-2.7mm} C}}
\nc{\Nat}{\mbox{\hspace{.04mm}\rule{0.2mm}{2.8mm}\hspace{-1.5mm} N}}

\nc{\spa}{\hspace{1 cm},\hspace{1 cm}}
\nc{\vs}{\vspace}
\nc{\NP}[1]{Nucl.\ Phys.\ {\bf #1}}
\nc{\PL}[1]{Phys.\ Lett.\ {\bf #1}}
\nc{\CMP}[1]{Commun.\ Math.\ Phys.\ {\bf #1}}
\nc{\PR}[1]{Phys.\ Rev.\ {\bf #1}}
\nc{\PRL}[1]{Phys.\ Rev.\ Lett.\ {\bf #1}}
\nc{\PTP}[1]{Prog.\ Theor.\ Phys.\ {\bf #1}}
\nc{\PTPS}[1]{Prog.\ Theor.\ Phys.\ Suppl.\ {\bf #1}}
\nc{\MPL}[1]{Mod.\ Phys.\ Lett.\ {\bf #1}}
\nc{\IJMP}[1]{Int.\ Jour.\ Mod.\ Phys.\ {\bf #1}}
\nc{\IM}[1]{Invent.\ Math.\ {\bf #1}}
\nc{\SJNP}[1]{Sov. J. Nucl. Phys.\ {\bf #1}}

\begin{document}

\topmargin -5mm
\oddsidemargin 5mm

\begin{titlepage}
\setcounter{page}{0}
\begin{flushright}
March 1998
\end{flushright}

\vs{8mm}
\begin{center}
{\Large Two-point Functions in Affine $SL(N)$ Current Algebra}

\vs{8mm}
{\large J{\o}rgen Rasmussen}\footnote{e-mail address: 
jorgen@celfi.phys.univ-tours.fr}\\[.2cm]
{\em Laboratoire de Math\'{e}mathiques et Physique Th\'{e}orique,}\\
{\em Universit\'{e} de Tours, Parc de Grandmont, F-37200 Tours, France}

\end{center}

\vs{8mm}
\centerline{{\bf{Abstract}}}
\noindent
In this letter the explicit form of general two-point functions in
affine $SL(N)$ current algebra is provided for all representations,
integrable or non-integrable.
The weight of the conjugate field to a primary field of arbitrary weight
is immediately read off.\\[.4cm]

\end{titlepage}
\newpage
\renewcommand{\thefootnote}{\arabic{footnote}}
\setcounter{footnote}{0}

\section{Introduction}

Two-point functions are the simplest non-trivial correlators one may
consider in (extended) conformal field theory. Nevertheless, results in
the case of general representations of affine current algebras
are still lacking, except for $SL(2)$ where invariance under
(loop) projective transformations immediately produces the result.

The objective of the present letter is to work out explicitly the 
two-point functions in affine $SL(N)$ current algebra for all 
representations, integrable or non-integrable. The construction is
based on the differential operator realization of simple Lie algebras
provided in \cite{PRY4} and on well-known results for fundamental
representations.

Besides providing us with new insight in the general structure of
conformal field theory based on affine current algebra, a motivation for 
studying two-point functions in affine current algebra is found in the
wish to understand how to generalize to higher groups
the proposal by Furlan, Ganchev, Paunov and Petkova
\cite{FGPP} for how Hamiltonian reduction of affine $SL(2)$
current algebra works at the level of correlators. A simple proof
of the proposal in that case
is presented in \cite{PRY2} based on the work \cite{PRY1}
on correlators for degenerate (in particular admissible)
representations in affine $SL(2)$ current algebra. Explicit knowledge
on two-point functions may be seen as a first step in the direction
of understanding that generalization.

Furthermore, an immediate application of knowing the two-point functions
is to determine the weight
of the conjugate (primary) field to a primary field of an arbitrary weight. 
This result is of importance since conjugate (or contragredient)
representations are very useful in many respects.
For non-integrable representations such weights are in general not known.

The remaining part of this letter is organized as follows. In Section 2
we review our differential operator realization \cite{PRY4} while
fixing the notation. In Section 3 the construction of two-point
functions in affine $SL(N)$ current algebra
is provided and the conjugate weights are derived.

\section{Notation}

Let {\bf g} be a simple Lie algebra of rank $r$.
{\bf h} is a Cartan subalgebra of {\bf g}. The set of (positive) roots
is denoted ($\Delta_+$) $\Delta$ and the simple roots are 
written $\al_i,\ i=1,...,r$. $\al^\vee = 2\al/\al^2$ is the root 
dual to $\al$. Using the triangular decomposition 
\ben
 \mbox{{\bf g}}=\mbox{{\bf g}}_-\oplus\mbox{{\bf h}}\oplus\mbox{{\bf g}}_+
\een
the raising and lowering operators are denoted $e_\al\in$ {\bf g}$_+$ and
$f_\al\in$ {\bf g}$_-$, respectively, with $\al\in\Delta_+$, and 
$h_i\in$ {\bf h} are the Cartan operators. 
In the Cartan-Weyl basis we have
\ben
 [h_i,e_\al]=(\al_i^\vee,\al)e_\al\spa [h_i,f_\al]=
  -(\al_i^\vee,\al)f_\al
\label{CW}
\een
and
\ben
 \left[e_\al,f_\al\right]=h_\al=G^{ij}(\al_i^\vee,\al^\vee)h_j
\een
where the metric $G_{ij}$ is related to the Cartan matrix $A_{ij}$ as
\ben
 A_{ij}=\al_i^\vee\cdot\al_j=(\al_i^\vee,\al_j)=
  G_{ij}\al_j^2/2
\een
The Dynkin labels $\La_k$ of the weight $\La$ are defined by
\ben
 \La=\La_k\La^{k}\spa \La_k=(\al_k^\vee,\Lambda)
\label{Dynkin}
\een
where $\left\{\La^{k}\right\}_{k=1,...,r}$ is the set of fundamental
weights satisfying
\ben
 (\al_i^\vee,\La^{k})=\delta_i^k
\een 
Elements in $\mbox{\bf g}_+$ may be parameterized using ``triangular 
coordinates" denoted by $x^\al$, one for each positive root, thus we 
write general Lie algebra elements in $\mbox{\bf g}_+$ as
\ben
 g_+(x)=x^\al e_\al \in \mbox{\bf g}_+
\label{gplus}
\een
We will understand ``properly" repeated root indices as in (\ref{gplus})
to be summed over the {\em positive} roots. Repeated Cartan indices as in
(\ref{Dynkin}) are also summed over.
The matrix representation $C(x)$ of $g_+(x)$ in the adjoint 
representation is defined by
\ben
 C_a^b(x)=-x^\beta {f_{\beta a}}^b
\label{cadj}
\een

Now, a differential operator realization $\left\{\tilde{J}_a(x,\pa,\Lambda)
\right\}$ 
of the simple Lie algebra {\bf g} generated by $\left\{j_a\right\}$
is found to be \cite{PRY4}
\bea
\tilde{E}_\al(x,\pa)&=&\vab(x)\db\nn
\tilde{H}_i(x,\pa,\Lambda)&=&\vib(x)\db+\Lambda_i\nn
\tilde{F}_\al(x,\pa,\Lambda)&=&\vmab(x)\db+\paj(x)\Lambda_j
\label{defVP}
\eea
where 
\bea
 \vab(x)&=&\left[B(C(x))\right]_\al^\beta\nn
 \vib(x)&=&-\left[C(x)\right]_i^\beta \nn
 \vmab(x)&=&\left[e^{-C(x)}\right]_{-\al}^\g\left[B(-C(x))\right]_\g^\beta\nn
 \paj(x)&=&\left[e^{-C(x)}\right]_{-\al}^j 
\label{VPQ}
\eea
$B$ is the generating function for the Bernoulli numbers
\ben
  B(u)=\frac{u}{e^u-1}=\sum_{n\geq 0}\frac{B_n}{n!}u^n\nn
\label{Ber}
\een
whereas $\pa_\beta$ denotes partial differentiation wrt $x^\beta$.
Closely related to this differential operator realization is the equivalent
one $\left\{J_a(x,\pa,\Lambda)\right\}$ given by
\bea
 E_\al(x,\pa,\La)&=&-\tilde{F}_\al(x,\pa,\La)\nn
 F_\al(x,\pa,\La)&=&-\tilde{E}_\al(x,\pa,\La)\nn
 H_i(x,\pa,\La)&=&-\tilde{H}_i(x,\pa,\La)
\label{tilde}
\eea
The matrix functions (\ref{VPQ}) are defined in terms of universal
power series expansions, valid for any Lie algebra, but ones that truncate 
giving rise to finite polynomials of which the explicit forms depend on the
Lie algebra in question.

In the case of $SL(N)$ the roots may be represented as 
\ben
 \al_{ij}={\bf e}_i-{\bf e}_j
\een
where $\left\{ {\bf e}_i\right\}$ is an orthonormal basis for the 
$N$-dimensional Euclidean space. The rank of $SL(N)$ is $r=N-1$.
The structure coefficients are given by
\ben
 [e_{ij},e_{kl}]=\delta_{jk}e_{il}-\delta_{li}e_{kj}
\een
supplemented by the symmetries
\bea
 {f_{-\al,-\beta}}^{-\g}&=&-{f_{\al,\beta}}^\g\nn
 {f_{\al,\beta}}^{\al+\beta}&=&
  {f_{\beta,-(\al+\beta)}}^{-\al}=
  {f_{-(\al+\beta),\al}}^{-\beta}
\label{symmf}
\eea
Here we have introduced the abbreviations
\ben
 e_{ij}=e_{\al_{ij}}\spa f_{ij}=f_{\al_{ij}}
\een

\subsection{Affine Current Algebra}

Associated to a Lie algebra is an affine Lie algebra characterized by the
central extension
$k$, and associated to an affine Lie algebra is an affine current
algebra whose generators are conformal spin one fields and have amongst
themselves the operator product expansion
\ben
 J_a(z)J_b(w)=\frac{\kappa_{ab}k}{(z-w)^2}+\frac{\fabc J_c(w)}{z-w}
\label{JaJb}
\een
where regular terms have been omitted.
$\kappa_{ab}$ is the Cartan-Killing form of the underlying Lie algebra.

It is convenient to collect the traditional multiplet of primary fields
in an affine current algebra (which generically is infinite for
non-integrable representations) in a generating function for that
\cite{FGPP,PRY1,PRY4}, namely the primary field 
$\phi_\La(w,x)$ which must satisfy
\bea
 J_a(z)\phi_\La(w,x)&=&\frac{-J_a(x,\pa,\La)}{z-w}\phi_\La(w,x)\nn
 T(z)\phi_\La(w,x)&=&\frac{\Delta(\phi_\La)}{(z-w)^2}\phi_\La(w,x)
  +\frac{1}{z-w}\pa\phi_\La(w,x)
\label{primdef}
\eea
Here $J_a(z)$ and $T(z)$ are the affine currents and the energy-momentum
tensor, respectively, whereas $J_a(x,\pa,\La)$ are the 
differential operator realizations. $\D(\phi_\La)$ denotes the conformal
dimension of $\phi_\La$.
The explicit construction of primary fields for general simple Lie algebra
and arbitrary representation is provided in \cite{PRY4}. 

An affine transformation of a primary field is given by
\bea
 \delta_\epsilon\phi_\La(w,x)&=&\oint_w\frac{dz}{2\pi i}
  \epsilon^a(z)J_a(z)\phi_\La(w,x)\nn
 &=&\left\{\epsilon^{-\al}(w)V_\al^\beta(x)\pa_\beta
  +\epsilon^i(w)\left(V_i^\beta(x)
  \pa_\beta+\La_i\right)\right.\nn
  &+&\left.\epsilon^\al(w)\left(V_{-\al}^\beta(x)\pa_\beta+
  P_{-\al}^i(x)\La_i\right)\right\}\phi_\La(w,x)
\label{Ward}
\eea
and is parameterized by the $d$ ($d$ is the dimension of the
underlying Lie algebra) independent infinitesimal functions $\epsilon^a(z)$. 

\section{Two-point Functions}

Let $W_2(z,w;x,y;\La,\La')$ denote a general two-point function of two 
primary fields $\phi_\La(z,x)$ and $\phi_{\La'}(w,y)$. From the conformal 
Ward identities or projective invariance the well-known conformal
property of the two-point function is found to be
\ben
 W_2(z,w;x,y;\La,\La')=\frac{\delta_{\D(\phi_\La),\D(\phi_{\La'})}}{(z-w)^{
 \D(\phi_\La)+\D(\phi_{\La'})}}W_2(x,y;\La,\La')
\een
The affine Ward identity 
\ben 
 \delta_\epsilon W_2(z,w;x,y;\La,\La')=
 \langle\delta_\epsilon\phi_\La(z,x)\phi_{\La'}(w,y)\rangle
 +\langle\phi_\La(z,x)\delta_\epsilon\phi_{\La'}(w,y)\rangle=0
\een
may be recast (using (\ref{Ward}))
into the following set of $d$ partial differential equations
\ben
 \left(\tilde{J}_a(x,\pa,\La)+\tilde{J}_a(y,\pa,\La')\right)
  W_2(x,y;\La,\La')=0
\label{ddiff}
\een
It is easily verified that only the $2r$ equations for $a=\pm\al_i$
are independent. By induction, this simply
follows from the fact that $\{\tilde{J}_a\}$
is a differential operator realization of a Lie algebra.
It is the general solution to the equations (\ref{ddiff})
that we shall provide in the case of $SL(N)$. 

The starting assumption, which we believe also to be
true for more general groups, is that the affine part of $W_2$ may be 
expressed as a product of $r$ monomials
\ben
 W_2(x,y;\La,\La')=\prod_{i=1}^r\left(R_i(x,y)\right)^{\mu_i(\La,\La')} 
\een 
where the functions $R_i(x,y)$ are common eigen-functions of the $d$ 
differential operators $V_a^\beta(x)\pa_{x^\beta}+V_a^\beta(y)\pa_{y^\beta}$.
Here we mean to include also vanishing eigen-values.
In the case of $SL(N)$ such eigen-functions may be obtained as follows.

First we review a well-known realization of the fundamental representations
of $SL(N)$ in terms of $N$ fermionic creation and annihilation
operators
\ben
 q_j^\dagger,\ q_j\spa j=1,...,N
\een
The $\binomial{N}{k}$ dimensional $k$'th
fundamental representation is provided by the set of states where $k$
fermionic creation operators act on the Fermi vacuum $\ket{0}$.
The highest weight vector of the $k$'th fundamental representation of
weight $\La^k$ is
\ben
 \ket{\La^k}=q_1^\dagger...q_k^\dagger\ket{0}
\een
whereas $e_{ij}$ and $f_{ij}$ are represented by
\bea
 e_{ij}&=&q_i^\dagger q_j\spa i<j\nn
 f_{ij}&=&q_j^\dagger q_i\spa i<j
\eea
A basis with a minimal set of lowering operators is then easily seen
to be 
\bea
 &&\ket{\La^k}\nn
 &&f_{ij}\ket{\La^k}\spa i\leq k<j\leq N\nn
 &&\vdots\nn
 &&f_{i_{p_n(1)}j_1}...f_{i_{p_n(n)}j_n}\ket{\La^k}\spa
  i_1<...<i_n\leq k<j_1<...<j_n\leq N\nn
 &&\vdots
\label{states}
\eea
where $p_n$ is a permutation operator in $n$ variables.
The simplest possible choice is of course the one where all these
permutations are identities. For notational reasons we will
stick to that case in the following, and since we only need one
basis this choice does not spoil the generality of our construction.
Furthermore, it is convenient to renormalize this basis in order that all
$\binomial{N}{k}$ basis elements
\ben
 q_{l_1}^\dagger...q_{l_k}^\dagger\ket{0}\spa 1\leq l_1<...<l_k\leq N
\een
appear with positive sign.
The corresponding (renormalized) polynomials are then
\bea
 b_0(x,\La^k)&=&1\nn
 b_1(x,\La^k;i,j)&=&(-1)^{k-i}\tilde{F}_{ij}(x,\pa,\La^k)1
  =(-1)^{k-i}P_{-\al_{ij}}^k(x)\ \ ,\ \ i\leq k<j\leq N\nn
 &\vdots&\nn
 b_n(x,\La^k;\{i_l\},\{j_l\})&=&(-1)^{n(k-(n-1)/2)-\sum_{l=1}^ni_l}
   \tilde{F}_{i_1j_1}(x,\pa,\La^k)...\tilde{F}_{i_{n-1}j_{n-1}}
(x,\pa,\La^k)\nn
 &\cdot&P_{-\al_{i_nj_n}}^k(x)\ \ ,\ \
   i_1<...<i_n\leq k<j_1<...<j_n\leq N\nn
 &\vdots&
\label{pol}
\eea

Our proposal for the fundamental eigen-functions $R_k(x,y)$ is the following
\bea
 &&R_k(x,y)=\nn
 &&\sum_{n,n'=0}^{min\{k,N-k\}}
   \tilde{\sum_{\{i_l\},\{j_l\};\{i'_{l'}\},\{j'_{l'}\}}}
  (-1)^{\sum_{l=1}^n(j_l-i_l)}b_n(x,\La^k;
   \{i_l\},\{j_l\})b_{n'}(y,\La^{N-k};\{i'_{l'}\},\{j'_{l'}\})\nn
\label{R}
\eea
where the second summation $\tilde{\sum}$ is restricted by
\ben
 \left(\{1,...,k\}\cup\{1,...,N-k\}\cup\{j_l\}\cup\{j'_{l'}\}\right)
 \setminus\left(\{i_l\}\cup\{i'_{l'}\}\right)=\{1,...,N\}
\label{cond}
\een
This restriction ensures that we only encounter terms where
the corresponding bi-product of states (\ref{states}) exactly "covers
the single state $q_1^\dagger...q_N^\dagger\ket{0}$". This may be 
illustrated as follows. Abbreviate a term in the summation by
$b_n(x)b_{n'}(y)$ where the corresponding states are
\bea
 b_n(x)&\sim&q_{m_1}^\dagger...q_{m_k}^\dagger\ket{0}\nn
 b_{n'}(y)&\sim&q_{m_1'}^\dagger...q_{m_{N-k}'}^\dagger
  \ket{0}
\eea
Such a pair of states may be represented by a diagram like in Figure 1,
where the "covering of $q_1^\dagger...q_N^\dagger\ket{0}$" is
transparent.\\[.2cm]
\noindent{\bf Figure 1}
\begin{center}
\begin{tabular}{cccccccccc}\hline
$\bullet$&$\bullet$&0&$\bullet$&...&$\bullet$&...&0&$\bullet$&0\\
 \hline
 0&0&$\bullet$&0&...&0&...&$\bullet$&0&$\bullet$\\
\hline
  1&2&3&4&...&$l$&...&.&.&$N$
\end{tabular}\\[.5cm]
\end{center}
A $\bullet$ in the upper line in place $l$
means that the corresponding $q_l^\dagger$
appears in the $x$-state and not in the $y$-state, and vice versa.
The combined summation in (\ref{R}) may be seen as a summation over
all such configurations, up to the sign factor which will be
accounted for in the following.

{}From the pictorial description it follows that
\bea
 \left(\tilde{E}_{ij}(x,\pa)+\tilde{E}_{ij}(y,\pa)\right) R_k(x,y)&=&0\nn
 \left(\tilde{F}_{ij}(x,\pa,\La^k)+\tilde{F}_{ij}(y,\pa,\La^{N-k})
  \right)R_k(x,y)&=&0
\label{EF}
\eea
Namely, a non-vanishing action of $e_{ij}(x)+e_{ij}(y)$
produces a configuration of the form depicted in Figure 2.\\[.2cm]
{\bf Figure 2}
\begin{center}
\begin{tabular}{ccccc}\hline
...&$\bullet$&...&0&...\\
 \hline
 ...&$\bullet$&...&0&...\\
\hline
  ...&$i$&...&$j$&...
\end{tabular}\\[.5cm]
\end{center}
This again may be obtained in two ways; from either of the two
configurations in Figure 3.\\[.2cm]
{\bf Figure 3}
\begin{center}
\begin{tabular}{ccccc}\hline
...&0&...&$\bullet$&...\\
 \hline
 ...&$\bullet$&...&0&...\\
\hline
  ...&$i$&...&$j$&...
\end{tabular}\\[.5cm]
\end{center}
\begin{center}
\begin{tabular}{ccccc}\hline
...&$\bullet$&...&0&...\\
 \hline
 ...&0&...&$\bullet$&...\\
\hline
  ...&$i$&...&$j$&...
\end{tabular}\\[.5cm]
\end{center}
However, due to the sign factor in (\ref{R}) they produce the configuration
in Figure 2 with opposite signs, hence the first equality
in (\ref{EF}). The second statement
in (\ref{EF}) is completely analogous. Note that generically the two
differential operators $\tilde{F}_{ij}(x,\pa,\La^k)$ 
and $\tilde{F}_{ij}(y,\pa,\La^{N-k})$ are defined
for two different fundamental representations.

We are now in a position to state the main result in this letter:\newpage
\noindent{\bf Proposition}\\
The two-point function of the primary fields $\phi_\La(z,x)$ and
$\phi_{\La'}(w,y)$ in affine $SL(N)$ current algebra is (up to an
irrelevant normalization constant) given by
\ben
 W_2(z,w;x,y;\La,\La')=\frac{\delta_{\D(\phi_\La),\D(\phi_{\La'})}}{(z-w)^{
 \D(\phi_\La)+\D(\phi_{\La'})}}
 \prod_{i=1}^r\left(R_i(x,y)\right)^{\mu_i(\La,\La')}
\een
where
\bea
 \mu_i(\La,\La')&=&\La_i\nn
  &=&\La'_{N-i}
\label{mu}
\eea
and $R_i(x,y)$ is given by (\ref{R}) and (\ref{cond}).\\[.2cm]
{\bf Proof}\\
As remarked above, we only need to consider the actions of 
$\tilde{E}_{\al_j}(x,\pa)+\tilde{E}_{\al_j}(y,\pa)$ and 
$\tilde{F}_{\al_j}(x,\pa,\La)+\tilde{F}_{\al_j}(y,\pa,\La')$ for $j=1,...,r$.
That the $r$ former operators satisfy (\ref{ddiff}) follows 
directly from (\ref{EF}). From (\ref{EF}) we also have 
\ben
 \left(V_{-\al}^\beta(x)\pa_{x^\beta}+V_{-\al}^\beta(y)\pa_{y^\beta}\right)
  R_i(x,y)=-\left(P_{-\al}^i(x)+P_{-\al}^{N-i}(y)\right)R_i(x,y)
\een
with no summation over $i$. This implies that
\bea
 &&\left(\tilde{F}_{\al_j}(x,\pa,\La)+\tilde{F}_{\al_j}(y,\pa,\La')\right)
   W_2(x,y;\La,\La')\nn
  &=&\left(\sum_{i=1}^r(\La_i-\mu_i(\La,\La'))P_{-\al_j}^i(x)
   +\sum_{i=1}^r(\La_{N-i}'-\mu_i(\La,\La'))
   P_{-\al_j}^{N-i}(y)\right)W_2(x,y;\La,\La')
\eea
{}from which we obtain (\ref{mu}).\\
$\Box$\\
{}From the condition on the pair $(\La,\La')$ in (\ref{mu}) it follows
immediately that the conjugate weight $\La^+$ to an arbitrary weight
$\La$, integrable or non-integrable, is given by
\ben
 \La^+=\sum_{k=1}^{r}\La_k^+\La^k=\sum_{k=1}^{r}\La_{N-k}\La^k
\een
This generalizes the well-known result for integrable representations where
$\La^+$ is given by minus the lowest weight
in the finite-dimensional highest weight representation of $\La$.

We hope to come back elsewhere with a discussion on two-point
functions (and conjugate weights) in affine current algebras
for more general groups and supergroups. In the latter cases one
may employ the recently obtained differential operator realizations 
of the underlying Lie superalgebras \cite{Ras2}.\\[.3cm]
{\bf Acknowledgment} \\[.2cm]
The author thanks Jens Schnittger for fruitful discussions and gratefully 
acknowledges the financial support from the Danish Natural Science
Research Council, contract no. 9700517.

\end{document}